\documentclass{article}

\usepackage{arxiv}

\usepackage{url}            
\usepackage[ruled,noline]{algorithm2e}
\usepackage{amsmath}
\usepackage{amssymb}
\usepackage{amsfonts}
\usepackage{caption}
\usepackage[colorinlistoftodos]{todonotes}
\usepackage{lineno}
\usepackage{mathtools}
\usepackage{float}
\usepackage{url}            
\usepackage{tabularx}

\usepackage{etoolbox}
\providetoggle{fullBackground}

\usepackage{parskip}  

\newcommand{\diag}{\text{diag}}

\newcommand{\prox}{\text{prox}}

\DeclareFontFamily{U}{wncy}{}
\DeclareFontShape{U}{wncy}{m}{n}{<->wncyr10}{}
\DeclareSymbolFont{mcy}{U}{wncy}{m}{n}
\DeclareMathSymbol{\comb}{\mathord}{mcy}{"58} 

\usepackage[T1]{fontenc}
\DeclareFontFamily{T1}{calligra}{}
\DeclareFontShape{T1}{calligra}{m}{n}{<->s*[1.44]callig15}{}
\DeclareMathAlphabet\mathcalligra   {T1}{calligra} {m} {n}

\usepackage{accents}
\newlength{\dhatheight}

\title{Calibrationless Multi-coil Magnetic Resonance Imaging with Compressed Sensing Using Physically Motivated Regularization}

\author{
  Nicholas Dwork\thanks{www.nicholasdwork.com, nicholas.dwork@cuanschutz.edu} \\
  Department of Radiology and Biomedical Imaging \\
  University of California in San Francisco
    \And
  Ethan M. I. Johnson \\
  Department of Biomedical Engineering \\
  Northwestern University
    \And
  Daniel O'Connor \\
  Department of Mathematics and Statistics \\
  University of San Francisco
    \And
  Jeremy W. Gordon \\
  Department of Radiology and Biomedical Imaging \\
  University of California in San Francisco
    \And
  Adam B. Kerr \\
  Center for Cognitive and Neurobiological Imaging \\
  Stanford University
    \And
  Corey A. Baron \\
  Robarts Research Institute \\
  The University of Western Ontario
    \And
  John M. Pauly \\
  Department of Electrical Engineering \\
  Stanford University
    \And
  Peder E. Z. Larson \\
  Department of Radiology and Biomedical Imaging \\
  University of California in San Francisco
}

\begin{document}
\maketitle

\begin{abstract}
  With the advent of multi-coil imaging and compressed sensing, a number of model based reconstruction algorithms have been created.  They incorporate a multitude of different regularization functions based on physics, observed phenomenology, and heuristics.  Moreover, several iterative methods exist that attempt to simultaneously estimate the sensitivity maps and the image.  In this manuscript, we present a generalization of several existing iterative model based algorithms.  We devise a calibrationless instance of this generalization that only incorporates regularization terms based on physics and the accepted compressed sensing phenomenology of sparsity in the wavelet domain.  We compare the results of the new amalgamated optimization problem with existing methods on both simulated and real datasets.  We show that the images reconstructed using the new method, entitled Multi-coil Compressed Sensing (MCCS), are of higher quality than existing methods in all cases studied.
\end{abstract}

\setcounter{footnote}{0}

\section{Introduction}

Multi-coil imaging (commonly called parallel imaging\footnote{The method for combining information from multiple coils to synthesize unknown k-space values was first described as \textit{parallel imaging} because it was thought that multiple k-space values were collected simultaneously \cite{sodickson1997simultaneous}.  This, however, implies that the net spin state of each isochromat is in two measurable states at the same time; this is not the case.  Additionally, multi-coil imaging requires $B_1^{-}$ sensitivity maps that have significant orthogonal components.  Thus, we feel that the \textit{parallel imaging} nomenclature is a misnomer and elect to call it \textit{multi-coil imaging} instead.}) and compressed sensing are two methods that have dramatically reduced the scan time required for Magnetic Resonance Imaging (MRI).  Multiple receive coils (antennas) improve signal-to-noise ratio (SNR) by using smaller coil elements closer to the subject and they can reduce scan time by exploiting additional spatial encoding \cite{bernstein2004handbook}.  Multiple coils are now routinely used in clinical MRI machines.

A popular and effective form of image reconstruction when using multiple coils is model-based reconstruction \cite{pruessmann1998coil,fessler2010model}.  With these algorithms, the sensitivity maps have historically been estimated first and the reconstructed image is determined in a subsequent step by solving some form of the following optimization problem \cite{pruessmann2001advances}:
\begin{equation}
  \underset{x}{\text{minimize}} \hspace{0.5em}
    \underbrace{
      (1/2) \, \left\| \boldsymbol{D}\, \boldsymbol{F}\, \boldsymbol{S} \, x - b \right\|_{\mathcal{N}^{-1}}^2
    }_{
      \text{Data consistency term}
    } + \lambda_x\,\mathcal{R}_x(x),
    \label{eq:modelBasedRecon}
\end{equation}
where $x\in\mathbb{C}^{MN}$ is the column-extended vector of the image $X\in\mathbb{C}^{M\times N}$, $\boldsymbol{D}=(D,D,\ldots,D)$ is a block diagonal matrix of $C$ block elements, $C$ is the number of receiver coils, $D$ is a diagonal matrix representing the sampling mask, $\boldsymbol{F}=\diag(F,F,\ldots,F)$ is a block diagonal matrix of $C$ block elements, $F$ represents the Discrete Fourier Transform (DFT), $\boldsymbol{S}=\left(S^{(1)}, S^{(2)}, \ldots, S^{(C)}\right)$ is a block column matrix, $S^{(j)}=\diag\left(s^{(j)}\right)$, $s^{(j)}$ is a column-extended vector of the $j^{\text{th}}$ sensitivity map, $b=(b^{(1)}, b^{(2)}, \ldots, b^{(C)})\in\mathbb{C}^{CK}$, $b^{(j)}\in\mathbb{C}^{K}$ is the vector of data collected from the $j^{\text{th}}$ receiver coil, and $\mathcal{R}_x:\mathbb{C}^{MN}\rightarrow\mathbb{R}$ is a regularization function with parameter $\lambda_x\geq 0$.  Here, and throughout the document, $(\cdot,\ldots,\cdot)$ denotes the concatenation of elements into a column and $\diag(\cdot)$ is the invertible function that converts a vector into a diagonal matrix with the input elements along the diagonal\footnote{In this manuscript, the range of $\diag$ is the set of diagonal matrices; thus it is onto and invertible.  The inverse function is not defined for matrices that are not diagonal.}.  The function $\|\cdot\|_{\mathcal{N}^{-1}}$ represents the norm of the inner product space $\mathbb{C}^{N}$ induced by the inverse of $\mathcal{N}=\text{Cov}(\eta,\eta)$, where $\text{Cov}$ denotes covariance, $\eta=(\eta^{(1)}, \eta^{(2)}, \ldots, \eta^{(C)} )\in\mathbb{C}^{CK}$, and $\eta^{(j)}$ is a vector of additive noise from the $j^{\text{th}}$ coil.
(Note that to ease readability, we have included a table with the definition of each term in Appendix \ref{sec:notationTable}.)

Compressed sensing incorporates an \textit{a priori} belief that there exists a transformation such that the result of specific linear transformation is sparse \cite{candes2008introduction,block2007undersampled,liu2009regularized,chun2015efficient,chun2017compressed}.  This belief takes the form of a specific regularization function in problem \eqref{eq:modelBasedRecon}, which reduces the total number of samples required for accurate image reconstruction.  The wavelet transform is often used as the sparsifying transformation:  $\mathcal{R}_x(x) = \| W\,x\|_1$; here, $W$ represents the wavelet transform and $\|\cdot\|_1$ represents the $\ell_1$ norm.  Model based compressed sensing algorithms include $L_1$-ESPIRiT \cite{uecker2014espirit}, SparseSENSE \cite{liu2008sparsesense}, and SENSE-LORAKS \cite{kim2017loraks}.

These algorithms assume that the estimated sensitivity maps are perfectly accurate; therefore, any inaccuracies in the sensitivity maps are absorbed into the image when satisfying the data consistency term of \eqref{eq:modelBasedRecon}.
Several methods attempt to alleviate this by simultaneously estimating both the sensitivity maps with the following optimization problem:
\begin{equation}
  \begin{aligned}
    \underset{s,x}{\text{minimize}}
      &\hspace{0.5em} (1/2) \left\| \boldsymbol{D} \, \boldsymbol{F} \, \boldsymbol{S} \, x
        - b \right\|_{\mathcal{N}^{-1}}^2 + \lambda_x \, \mathcal{R}_x(x) + \lambda_s \, \mathcal{R}_s(s) \\
    \text{subject to} & \hspace{0.5em} \text{constraints on } x \text{ and } s,
  \end{aligned}
  \label{eq:xsModelBasedRecon}
\end{equation}
where  $\mathcal{R}_s$ is a regularization function with regularization parameter $\lambda_s\geq 0$, and $s=\left( s^{(1)}, s^{(2)}, \ldots, s^{(C)} \right)$.
The significant difference is that the sensitivity maps and the image are jointly estimated from the data; thus, errors are distributed between the optimization variables.  This problem is not generally a convex optimization problem due to the multiplication of $\boldsymbol{S}$ and $x$.  

Methods that fall into the form of \eqref{eq:xsModelBasedRecon} include JSENSE \cite{ying2007joint}, iSENSE \cite{majumdar2012iterative} Sparse BLIP \cite{she2014sparse}, and NLINV \cite{uecker2008image}.
With JSENSE, $\mathcal{R}_x=0$, $\mathcal{R}_s=0$, and the coil sensitivity maps are constrained to satisfy low order polynomials \cite{ying2007joint}.
There are two versions of iSENSE \cite{majumdar2012iterative}.  With iSENSE-CS, $\mathcal{R}_x=\|W x\|_1 + \gamma\,\|\nabla x\|_{2,1}$ where $\gamma>0$ (which assumes sparsity of the wavelet transform and the gradient field) and $\mathcal{R}_s= \sum_{j=1}^C \| S^{(j)}\|_\ast$ (the sum of the nuclear norms of the diagonal sensitivity map matrices).  With iSENSE-NN, $\mathcal{R}_s$ is the same as iSENSE-CS but $\mathcal{R}_x=\|X\|_\ast$.  The Sparse BLIP method incorporates the same $\mathcal{R}_x$ as iSENSE-CS but also imposes total variation regularization on the sensitivity maps. With NLINV, $\mathcal{R}_x(x)=\|x\|_2^2$,  $\mathcal{R}_s(s)=$, and $\lambda_x=\lambda_s$.  The iSENSE-CS, iSENSE-NN, Sparse BLIP, and NLIVNV algorithms all have a data consistency term that is the $\ell_2$ norm squared; therefore, they implicitly assume that the noise from different coils is uncorrelated (i.e., $\mathcal{N}$ is a scaled identity matrix).  Where as JSENSE, iSENSE, and Sparse BLIP each solves \eqref{eq:xsModelBasedRecon} with an alternating minimization algorithm, NLINV solves the problem with a gradient descent approach.

The methods above incorporate several heuristic regularization terms which have proven themselves to be useful in some cases.  However, these terms have not been justified with fundamental physics.  For example, the regularization of the sensitivity map in iSENSE is equivalent to $\mathcal{R}_s=\|s\|_1$, which represents the \textit{a priori} belief that the values of the sensitivity maps are distributed according to a Laplacian with many values equal to $0$; this is almost certainly not the case.  The regularization of the image values in NLINV (also used in the extension ENLIVE \cite{holme2019enlive}) incorporate the \textit{a priori} belief that the magnitudes of its values are normally distributed; this is almost certainly not the case.
The methods' reliance on functions that are not physically motivated may prevent fully exploiting the totality of the information present in the data or they may constrain the reconstruction in inappropriate ways.

\section{Background}
\label{sec:background}

In this work, we provide a model-based multi-coil reconstruction algorithm of the form of \eqref{eq:xsModelBasedRecon} where we only include regularization terms and constraints that are motivated by fundamental physics (justifiable by Maxwell's equations) and the established phenomenology of sparsity in the wavelet domain.
Three separate aspects of electromagnetics are included in the optimization problem that is solved to reconstruct the image.
\begin{enumerate}
  \item In accordance with the Principle of Reciprocity, the sensitivity maps are bounded in magnitude \cite{hoult2000principle}.

  \item In clinical MRI, it is accurate to make the approximation that the wavelength emitted from the coils is large compared to the receiver coil size \cite{bankson2002simulation}.  With this assumption, it follows that the sensitivity maps can be estimated by the Biot-Savart law \cite{roemer1990nmr,dominguez2016intensity,hansen2019coil,ohliger202055mn}.  Thus, the vast majority of the energy in the Fourier transforms of the sensitivity maps is contained within a small bandwidth.

  \item A coil array generates sensitivity maps with a small number of significant eigenmodes \cite{king2010optimum,buehrer2007array}.  (This fact is used with success by coil compression algorithms \cite{zhang2013coil,buehrer2007array}.)  This indicates that the nuclear norm of a matrix comprised of columns of sensitivity map vectors is small.
  
\end{enumerate}
Each aspect listed will be incorporated into an instance of problem \eqref{eq:xsModelBasedRecon} as a regularization term or a constraint.

Let us consider the data consistency term.  Suppose that $x^\star$ and $s^\star$ were solutions to problem \eqref{eq:xsModelBasedRecon}.  Without any constraints imposed on the magnitudes of $x$ and $s$, then $x^\star a$ and $s^\star /a$ would also be an optimal solution for any positive scalar $a$.  Even so, $x^\star a$ would remain proportional to magnetic density.  However, the ambiguity could lead to numerical instabilities during the optimization, where one or the other of $x$ or $s$ could grow to exceed representable values.  In the case of JSENSE, for example, where $\mathcal{R}_s=0$, the $\mathcal{R}_x$ regularization term encourages the image values to be small.  Thus; JSENSE is not guaranteed to converge to a solution.  If JSENSE were permitted to run for a large number of iterations, this would lead to an explosion of the sensitivity values and a shrinking of the image values.  JSENSE prevents this with early stopping \cite{mahsereci2017early}, which omits an optimal solution.

To enable running an alternating minimization for enough iterations so that an optimal (though perhaps only locally optimal) solution is attained while not subjecting the result to the scaling ambiguity, we use the fact that the sensitivity maps are bounded.  The bound is not known due to ambiguities in the way the data was collected (e.g., unknown amplifier gains and scaling of the analog-to-digital converter used to digitize the received signals).  Consider $\iiint_{-\infty}^{\infty} \sigma^{(i)}\,x\,dV$, where $\sigma:\mathbb{R}^3\rightarrow\mathbb{C}$ is the sensitivity of the $i^{\text{th}}$ coil; the maximum magnitude of this value is attained when all of the phases of $\sigma^{(i)}\,x$ are aligned.  Ideally, this is the Fourier value of zero frequency, but may not be due to phase inhomogeneities of the image or the sensitivity maps.  As a surrogate for this maximum value, we use $\max(|b|)$ and normalize the data $|b|$ as follows: $b := b / \max(|b|)$.  (Here, $|b|$ is a vector of magnitudes for corresponding values in $b$.)  Then, we impose a bound of $1$ on the magnitude of each element of the sensitivity maps: $|s_i|\leq 1$ for all $i$.  Suppose this bound were imposed on JSENSE, where there is a regularization on the  not any regularization imposed on the sensitivity maps; then the value of $a$ 

Assuming there is no regularization on the sensitivity maps, the regularization terms of the image will still encourage the image values to be small.  The image values will be reduced until the maximum magnitude of the sensitivity maps becomes $1$.  With this bound, a large number of iterations of the optimization algorithm can be employed, permitting the attainment of an optimal solution.

The low bandwidth of the sensitivity maps implies that the energy in the high frequencies of $F s^{(i)}$ should be penalized.  An extreme version of this would be to impose a constraint so that the energy in all but the lowest frequencies must be $0$.  However, this assumption limits the possible realizations of the sensitivity maps.  In particular, not all sensitivity maps that satisfy the Biot-Savart law can be represented well with this assumption.  Instead, we elect to incorporate a regularization function similar to that of NLINV, which permits non-zero energy in the high frequencies but penalizes that energy in a regularization term \cite{uecker2008image,holme2019enlive}.


\section{Methods}
\label{sec:methods}

We propose to solve the following instance of \eqref{eq:xsModelBasedRecon} which incorporates the physical aspects described in section \ref{sec:background}:
\begin{equation}
  \begin{aligned}
    \underset{s,x}{\text{minimize}} &
      \hspace{0.5em} (1/2) \left\| \boldsymbol{D} \boldsymbol{F} \boldsymbol{S} x - b \right\|_{\mathcal{N}^{-1}}^2 +
      \lambda_x \, \|W\,x\|_1 + \lambda_s \, \| \mathcal{S} \|_\ast + 
        ( \tilde{ \lambda }_s / 2 ) \, \|\boldsymbol{D}_c \, \boldsymbol{F} \, s \|_2^2 \\
    \text{subject to} & \hspace{1em}  |s_i| \leq 1 \hspace{0.5em} \text{ for all } \hspace{0.5em} i.
  \end{aligned}
  \label{eq:mccsProb}
\end{equation}
Here, $\mathcal{S} = [ s^{(1)} \, s^{(2)} \,  \cdots \, s^{(C)} ] \in \mathbb{C}^{MN\times C}$.
The matrix $\boldsymbol{D}_c=\diag(D_c, D_c, \cdots, D_c)$ where $D_c$ is a diagonal mask (where all values along the diagonal are either $1$ or $0$) that isolates the high frequencies (those frequencies above the FWTM of the coil used).
The sparsifying operator $W$ is the discrete Daubechies-4 wavelet transform \cite{daubechies1992ten} with scales \cite{lustig2007sparse}.
The regularization functions are $\mathcal{R}_x(x) = \|W\,x\|_1$, and $\mathcal{R}_s(s)=\|\mathcal{S}\|_\ast + \tilde{ \lambda }_s / (2\lambda_s) \| \boldsymbol{D}_c \, \boldsymbol{F} \, s \|_2^2 $.  Simulations with the Biot-Savart law \cite{esin2017mri} for three different coil arrangements (representative of those used to collect the data analyzed in this manuscript) are used to estimate the cutoff frequency of the regularization term.  Specifically, the full-width tenth maximum (FWTM) frequency of the power spectral density is used as the cutoff frequency.

As with existing methods, we solve problem \eqref{eq:mccsProb} using an alternating minimization algorithm that iterates over 1) solving for the sensitivity maps using the current estimate of the image, and 2) solving for the image with the current estimate of the sensitivity maps.  The alternating minimization algorithm is guaranteed to linearly converge to a local optima \cite{both2022rate,shi2016primer}.  The algorithms used to solve each of the sub-problems are described below.

\subsection{Estimating the sensitivity maps}

We first describe the method for estimating the sensitivity maps given an estimate of the image.  Since the coils are (almost always) placed externally to the imaged subject and the sensitivities of the coils extend both inwards towards the subject and outwards away from the subject, the support of the sensitivity maps is necessarily much larger than that of the image.  Therefore, in order to prevent aliasing when estimating the sensitivity maps, one must increase the field of view of the estimates.  For the results presented in this work, we assumed that the field of view of the sensitivity maps was less than twice that of the field of view of the image:  $FOV_s < 2 \, FOV_x$.

Note that $S^{(c)}x = s^{(c)} \odot x = \boldsymbol{X} s^{(c)}$ where $\odot$ denotes the Hadamard (or point-wise) product and $\boldsymbol{X}~=~\diag(X,X,\ldots,X)$.  With this conversion, one can estimate the sensitivity maps by solving the following optimization problem:
\begin{equation}
  \begin{aligned}
    \underset{s}{\text{minimize}} &
      \hspace{0.5em} (1/2) \left\| \boldsymbol{D} \boldsymbol{F} \boldsymbol{X} s - b \right\|_{\mathcal{N}^{-1}}^2
        + \lambda_s \|\mathcal{S}\|_\ast + ( \tilde{ \lambda }_s / 2 ) \, \| \boldsymbol{D}_c \, \boldsymbol{F} \, s \|_2^2 \\
    \text{subject to} & \hspace{1em}  |s_i| \leq 1 \hspace{0.5em} \text{ for all } \hspace{0.5em} i.
  \end{aligned}
  \label{eq:mccs_findMaps}
\end{equation}

Let $\mathcal{N}^{-1} = \boldsymbol{L} \boldsymbol{L}^H$ be the Cholesky decomposition of the inverse noise covariance matrix, where $\boldsymbol{L}^H$ denotes the Hermitian (or conjugate transpose) of $\boldsymbol{L}$.  To solve the problem, we first convert the data consistency term into an $\ell_2$ norm squared as follows:
\begin{equation}
  \begin{aligned}
    \left\| \boldsymbol{D} \boldsymbol{F} \boldsymbol{X} s - b \right\|_{\mathcal{N}^{-1}}^2
      &= \left\| \boldsymbol{L}^H \boldsymbol{D} \boldsymbol{F} \boldsymbol{X} s - \boldsymbol{L}^H b \right\|_2^2 \\
      &= \left\| \boldsymbol{D} \boldsymbol{F} \boldsymbol{L}^H \boldsymbol{X} s - \boldsymbol{L}^H b \right\|_2^2.
  \end{aligned}
  \label{eq:l2DataConsistency}
\end{equation}
Here, we used the fact that $\boldsymbol{L}^H$ and $\boldsymbol{D} \boldsymbol{F}$ commute \cite{pruessmann2001advances}.  Let $P$ be a permutation matrix that reorders the elements from having all k-space samples of the same coil adjacent to each other to having all coil samples of the same k-space location adjacent to each other.  Then $\boldsymbol{L}=P^T \left( L \otimes I_K \right) P$, where $P^T$ denotes the transpose of $P$, $L\in\mathbb{C}^{C\times C}$, $\otimes$ denotes the Kronecker product, and $I_K$ is the identity matrix of size $K\times K$ \cite{pruessmann2001advances}.  This expression permits left multiplication by $\boldsymbol{L}^H$ without consuming much memory.

Incorporating \eqref{eq:l2DataConsistency} into \eqref{eq:mccs_findMaps} yields
\begin{equation}
\begin{aligned}
    \underset{s}{\text{minimize}} & \hspace{1em} 
      \left\| \boldsymbol{D} \boldsymbol{F} \boldsymbol{L}^H \boldsymbol{X} s - \boldsymbol{L}^H b \right\|_2^2
      + \lambda_s \| \boldsymbol{\mathcal{S}} \|_\ast + (\tilde{\lambda}_s / 2) \, \| \boldsymbol{D}_c \, \boldsymbol{F} \, \diag(s) \|_2^2 \\
    \text{subject to} & \hspace{1em} |s_i| \leq 1 \text{ for all } i,
    \label{eq:mccsProbDwork_findMapsL2}
  \end{aligned}
\end{equation}

We solve this problem with the Primal-Dual Hybrid Gradient algorithm \cite{chambolle2011first} (as detailed in Appendix \ref{sec:pdhgForSenseMaps}).  Note that $\boldsymbol{L}^H \boldsymbol{X}$ and $\boldsymbol{L}^H b$ can both be computed once and stored in memory prior to any iterations of an optimization algorithm to reduce computation \cite{pruessmann2001advances}.

\subsection{Estimating the image}

To reconstruct the image for a given set of sensitivity maps, one solves the following optimization problem:
\begin{equation*}
    \underset{x}{\text{minimize}}
      \hspace{0.5em} (1/2) \left\| \boldsymbol{D} \boldsymbol{F} \boldsymbol{L}^H \boldsymbol{S} x - \boldsymbol{L}^H b \right\|_2^2 +
      \lambda_x \, \|W\,x\|_1.
\end{equation*}

Using the same Cholesky decomposition of $\mathcal{N}^{-1}$ as described above, this problem becomes
\begin{equation}
  \underset{x}{\text{minimize}} \hspace{1em} 
    (1/2) \left\| \boldsymbol{D} \boldsymbol{F} \boldsymbol{L}^H \boldsymbol{S} x - \boldsymbol{L}^H b \right\|_2^2
    + \lambda_x \, \| W \, x \|_1.
  \label{eq:mccs_findImage}
\end{equation}

Since $W$ is orthogonal, $\mathcal{R}_x(x) = \|W\,x\|_1$ has a simple proximal operator.  We solve this problem with the proximal optimal gradient method (POGM) \cite{fessler2019opt,kim2018adaptive}.  Again, $\boldsymbol{L}^H \boldsymbol{S}$ and $\boldsymbol{L}^H b$ can be computed once and stored in memory to reduce computation time.

\subsection{Multi-coil Compressed Sensing}

The complete Multi-coil Compressed Sensing (MCCS) algorithm is presented in Alg. \ref{alg:mccs}.
\begin{algorithm}[ht]
    \protect\caption{Multi-coil Compressed Sensing (MCCS)}
    \label{alg:mccs}

    \textbf{Inputs:}  $b$, $\lambda_x$, $\lambda_s$, $\tilde{ \lambda }_s$, $Z$

    \textbf{Initialize:} Initialize $s_{(0)}$ and $x_{(0)}$.

    $b := b / \max(|b|)$

    \textbf{For} $\zeta = 1, 2, \ldots, Z$

    \hspace{1em}  Determine $s_{(\zeta)}$ using $x_{(\zeta-1)}$ by solving \eqref{eq:mccs_findMaps} with PDHG.
    
    \hspace{2em}  The problem is initialized with $s_{(\zeta-1)}$.

    \hspace{1em}  Determine $x_{(\zeta)}$ using $s_{(\zeta)}$ by solving \eqref{eq:mccs_findImage} with POGM.
    
    \hspace{2em}  The problem is initialized with $x_{(\zeta-1)}$.

    \textbf{End For}

    $X=\diag(x{(Z)})$

    \textbf{Outputs: } $X$

\end{algorithm}

For the results presented in this manuscript, the sensitivity map for coil $c$ is initialized to the zero filled reconstruction divided by the root-sum-of-squares reconstruction and the image $x^{(0)}$ is initialized to the reconstruction of Roemer et al. \cite{roemer1990nmr}.

\section{Experiments}
\label{sec:experiments}

We compare results of solving \eqref{eq:mccsProb} using Alg. \ref{alg:mccs} to  SAKE+L1 ESPIRiT \cite{shin2014calibrationless,uecker2014espirit} and SENSE-LORAKS \cite{kim2017loraks} for simulated data of the brain as well as data of the knee and ankle.  Unless otherwise specified, the default parameters provided with the SAKE+L1 ESPIRiT software were used.  For SAKE, a kernel size of $6$, a window threshold of $1.8$, and $100$ iterations were used.  For L1 ESPIRiT, an eigenvalue threshold of $0.9$ and a regularization parameter of $2.5\cdot 10^{-3}$ were used.  SENSE-LORAKS does not have any parameters.

All data was collected on a 3DFT Cartesian trajectory.  For the real data, a one-dimensional inverse Discrete Fourier Transform was applied along the readout direction, which placed the data into the $(k_x,k_y,z)$ hybrid space.  We then isolated a single slice for further processing.

For MCCS, the noise correlation matrix $\mathcal{N}$ can be determined either with 1) a scan without any excitation so that all of the signals received are only noise, or 2) a region of the images without any sample (containing predominantly noise).  For the real data, we chose the latter technique (estimating noise statistics in regions of the image without any sample).

For SAKE+L1 ESPIRiT and MCCS, reconstructions were generated for a set of regularization parameters.  For SAKE+L1 ESPIRiT, the regularization parameters used were $2.5\cdot 10^{-5}, 2.5\cdot 10^{-4}, 2.5\cdot 10^{-3}, 2.5\cdot 10^{-2}, 2.5\cdot 10^{-1}$.  (Note that the default parameter value supplied with SAKE+L1 ESPIRiT is $2.5\cdot 10^{-3}$.)  Reconstructions were generated for MCCS with all combinations of $\lambda_x~\in~\left\{ 10^{-14}, 10^{-13}, 10^{-12}, 10^{-11}, 10^{-10}, 10^{-9},10^{-8},10^{-7},\right.$ $\left.10^{-6} \right\}$ and $\lambda_s~\in~\left\{ 10^{-10}, 10^{-9}, 10^{-8},10^{-7},10^{-6},10^{-5}, 10^{-4},\right.$ $\left. 10^{-3},10^{-2}, 10^{-1}, 1 \right\}$.

For the results presents, the number of iterations of the MCCS algorithm specified in Alg. \ref{alg:mccs} was $Z=50$.  The number of iterations for the PDHG method used to solve \eqref{eq:mccs_findMaps} was $90$.  The number of iterations for POGM used to solve \eqref{eq:mccs_findImage} was $30$.  Note that if $k$ is the iteration number, then the convergence rate of POGM is $\mathcal{O}(1/k^2)$ while the convergence rate of PDHG is $\mathcal{O}(1/k)$, which is why fewer iterations were required for POGM.

\subsection{Simulation}

\begin{figure}[ht]
  \centering{}
  \includegraphics[width=0.95\linewidth]{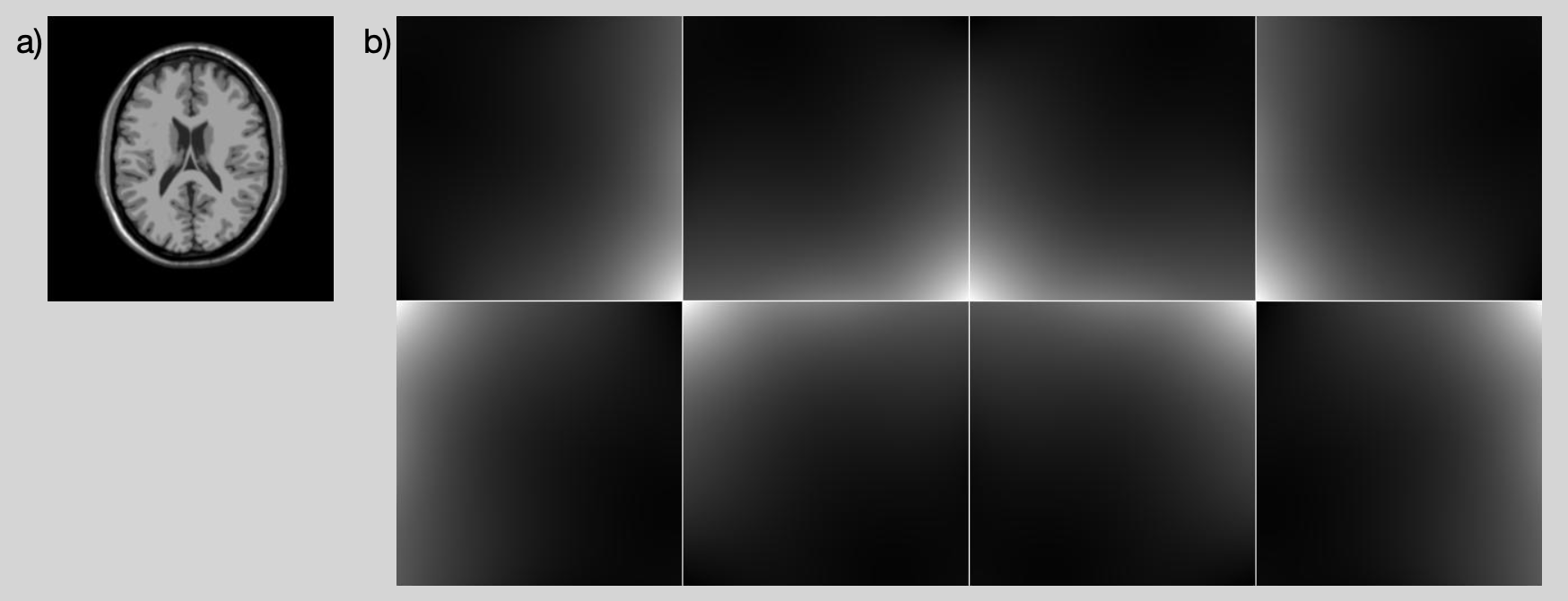}
  \caption{ \label{fig:simData}  Simulation of a eight channel receiver system.  a) Shows the original image and b) shows the coil sensitivity maps.}
\end{figure}

Multi-coil data were generated with simulations of an axial slice of a brain created with the BrainWeb simulation software \cite{kwan1999mri,cocosco1997brainweb}.  Eight rectangular coils were used in the simulation; they were evenly spaced around the brain with a distance of $0.5$ meters between opposite coils.  The Biot-Savart law was used to simulate the sensitivity maps for each coil \cite{esin2017mri}.  Coil coupling was simulated by constructing the sensitivity matrix $\mathcal{S}$ and projecting it onto the closest matrix (in a Frobenius sense) of rank $5$.
Figure \ref{fig:simData} shows the simulated image and the sensitivity maps of the simulated data.

For the numerical phantom, the mutual information metric is used for evaluation of the reconstructions of the phantom.  Mutual information was chosen because differences in scaling of the intensity between images is irrelevant.  The regularization parameters that yielded the highest mutual information between the reconstructed image and the simulated truth were used to generate the output.

\subsection{Real Data}

Publicly available data of a knee was acquired from \url{mridata.org} \cite{ong2018mridata}.
Data of the ankle was acquired from a healthy volunteer on a 3 Tesla scanner (MR750, GE Healthcare) equipped with clinical imaging gradients (5 G/cm maximum strength, 20 G/cm/ms maximum slew-rate).
The MR data of the ankle was gathered with Institutional Review Board (IRB) approval and Health Insurance Portability and Accountability Act (HIPAA) compliance.  Informed consent was obtained from the participant included in the study.
All data used were collected with $8$ channel coil arrays.
All data collected was fully sampled and then retrospectively downsampled for processing.

Knee data were acquired with an 8-channel extremity coil using a Spin Echo acquisition in an axial orientation.  Scan parameters were FOV = $16.0 \times 16.0 \times 15.4$ mm$^3$, matrix size = $320 \times 320 \times 256$ with $2 \times 2 \times 0.6$ mm$^3$ resolution, and TR / TE = 1550 / 25 ms.
Ankle data were acquired with an 8-channel foot and ankle coil using a 3D SPGR acquisition in a sagittal orientation. Scan parameters were FOV = $25.6 \times 25.6 \times 10.4$ cm$^3$, matrix size = $256 \times 256 \times 104$ with 1 mm$^3$ isotropic resolution, and TR / TE = $14.0$ / $3.0$ ms.


As there is no truth for the real data, we observed the outputs from all regularization parameters and present the one that appeared best qualitatively.

\section{Results}
\label{sec:results}

Results were generated for comparison using SENSE-LORAKS \cite{kim2017loraks}, SAKE+L1 ESPIRiT \cite{shin2014calibrationless,uecker2014espirit}, and MCCS.

\begin{figure}[ht]
  \centering{}
  \includegraphics[width=0.6\linewidth]{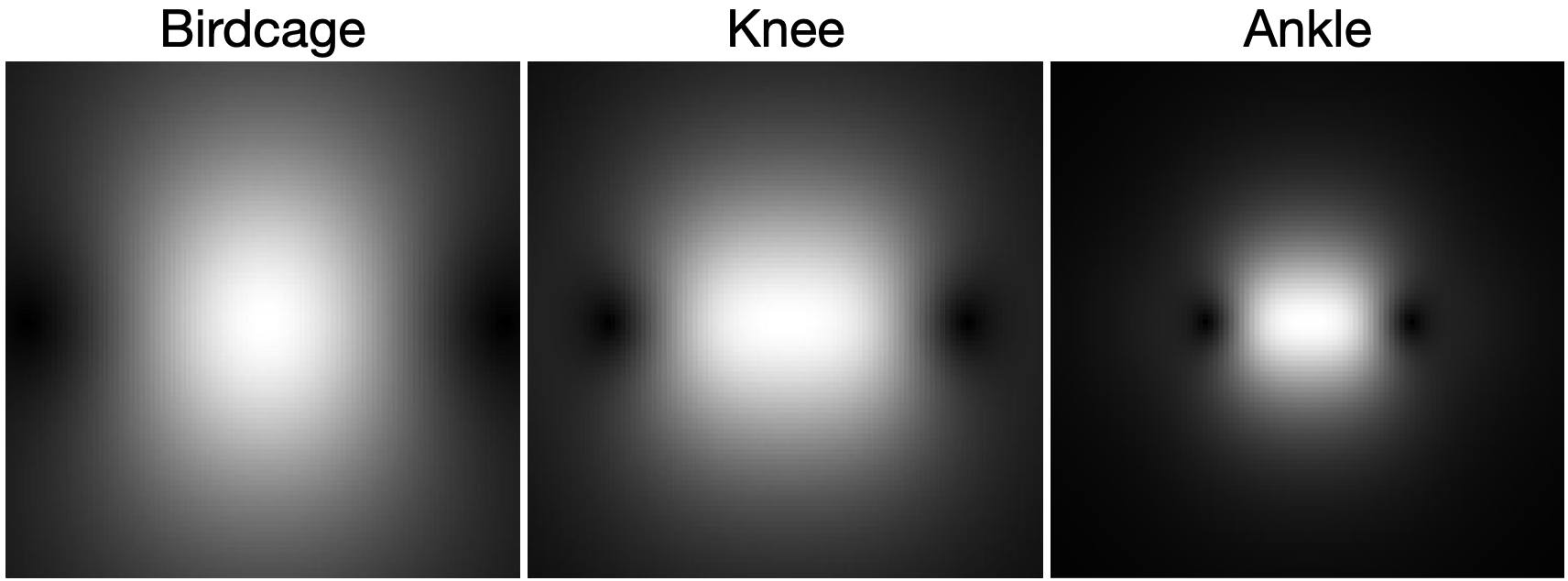}
  \caption{ \label{fig:coilSenseMaps}  Sensitivity maps simulated using the Biot-Savart law for (left) a birdcage coil of dimensions $12$ cm by $8$ cm, (center) a surface coil of dimensions $7$ cm by $7$ cm, and (right) a foot and ankle coil of size $4$ cm by $4$ cm.  The coils are located in the center of the image, lay horizontally, and are centered so that they extend into and out of the page. The images are $100\times 100$ and each pixel represents a $1.5\times 1.5$ mm$^2$ area. }
\end{figure}

Simulations with the Biot-Savart law \cite{esin2017mri} for three different coil arrangements (representative of those used to collect the data analyzed in this manuscript) are used to estimate the cutoff frequency of the regularization term (Fig. \ref{fig:coilSenseMaps}).  The coils simulated are a single coil from a birdcage coil \cite{giovannetti2002fast}, a single coil from a surface coil array \cite{hardy200632}, and a single coil from an ankle and foot coil array \cite{reach2007compartments}.  The birdcage, surface, and ankle coils have FWTM values of $3$, $5$, and $9$ cycles per meter, respectively.  These are the cutoff frequencies used for the corresponding results presented in this manuscript.

When imaging the numerical phantom; the regularization parameter that yielded the largest mutual information with the reference image was reported.    Recall that the Biot-Savart simulation found a FWTM frequency of $3$ cycles per meter for this coils used in this dataset.  The set of regularization parameters that yielded the highest mutual information with the reference image for MCCS were $(\lambda_x, \lambda_s) = (10^{-7},10^{-6})$.  The regularization parameter that yielded the highest mutual information with the reference image for SAKE+L1 ESPIRiT was $2.5\cdot 10^{-2}$.  There aren't any parameters for the SENSE-LORAKS algorithm.  As shown in table \ref{tbl:reconMIs}, MCCS attained the highest mutual information with the reference image.

\begin{table}[ht]
  \centering
\begin{tabular}{ |c|c|c|c|c|  }
    \hline
    SENSE-LORAKS & SAKE+L1 ESPIRiT & MCCS \\
    \hline
    $1.21$ & $1.54$ & $2.24$ \\
    \hline
  \end{tabular}
  \caption{Mutual Information between reference image and undersampled reconstruction }
  \label{tbl:reconMIs}
\end{table}

\begin{figure*}
  \centering{}
  \includegraphics[width=0.8\linewidth]{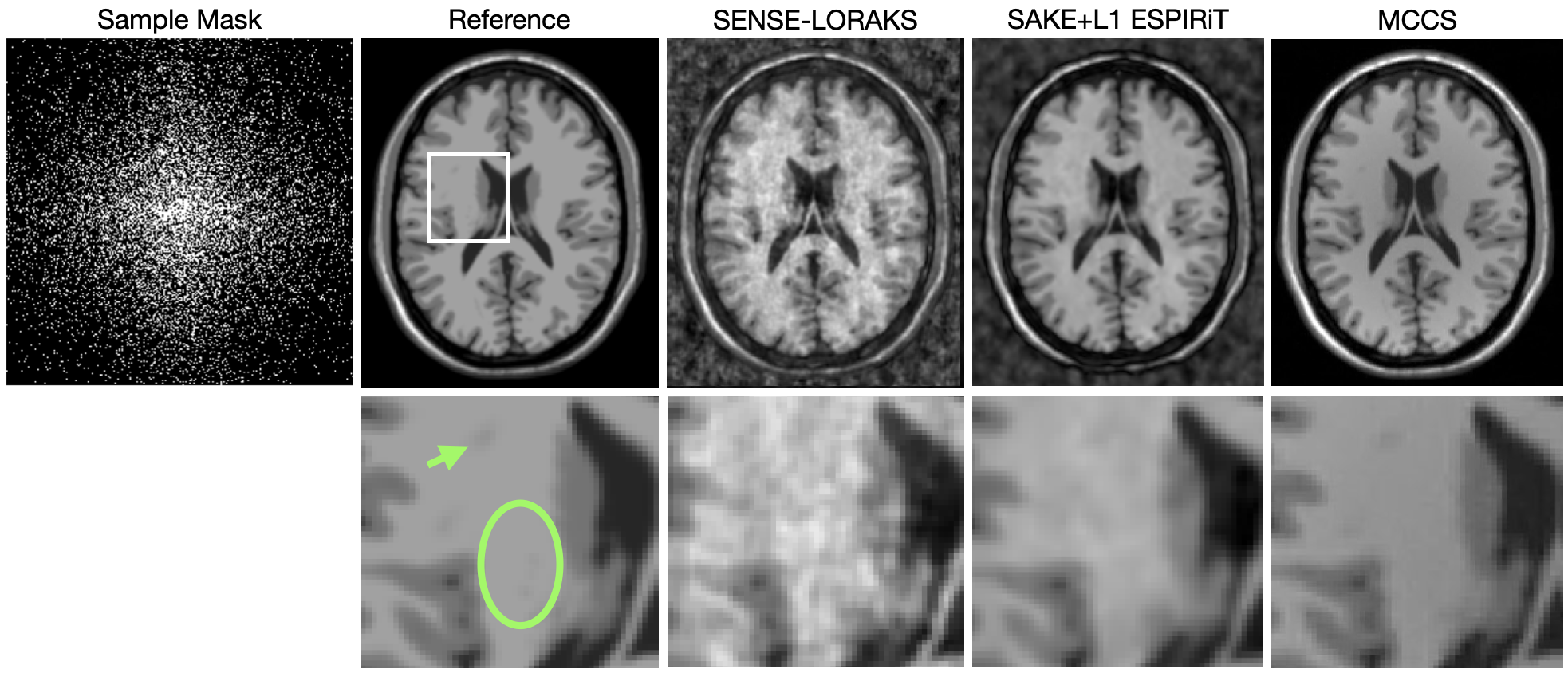}
  \captionsetup{width=.95\linewidth}
  \caption{ \label{fig:phantomRecons20}  The left column shows the sample mask used for this data (white points represent samples that were collected).  The top row shows the full field of view, the bottom row is zoomed into the white box overlaid on top of the reference image.  The third, fourth, and fifth columns present reconstructions using SENSE-LORAKS, SAKE+L1 ESPIRiT, and MCCS with the sample mask shown, which is 20\% of the data require for full sampling (that which satisfies the Nyquist-Shannon theorem).  The green arrow points to a detail in the brain that is not apparent in the SENSE-LORAKS or SAKE+L1 ESPIRiT reconstruction but is visible in the MCCS reconstruction.  The green ellipse encloses a region of high noise in the SENSE-LORAKS reconstruction but is not as noisy in the MCCS reconstruction. }
\end{figure*}

Figure \ref{fig:phantomRecons20} shows reconstructions of the brain simulations.  MCCS is able to retain detail better than SAKE+L1 ESPIRiT; an example is depicted by the point indicated with the green arrow.  MCCS is able to reduce noise better than SENSE LORAKS as shown in the region outside of the brain and the area enclosed by the green oval.

\begin{figure*}
  \centering{}
  \includegraphics[width=0.8\linewidth]{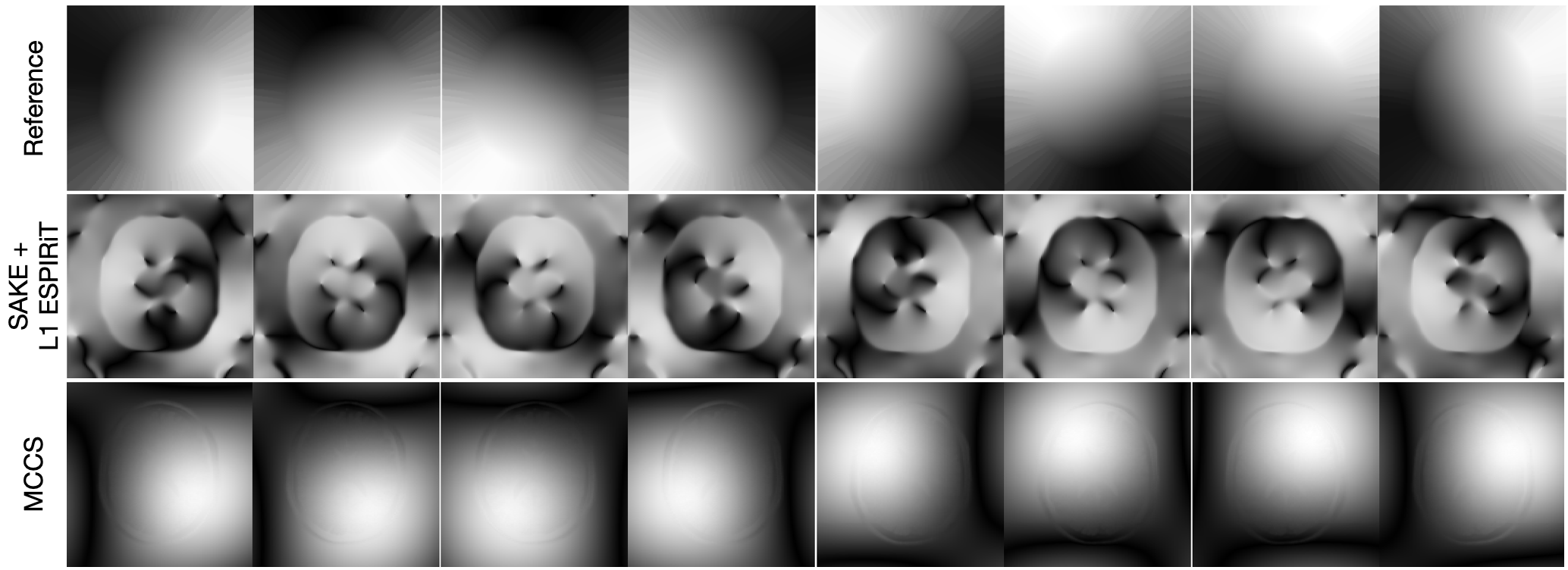}
  \captionsetup{width=.95\linewidth}
  \caption{ \label{fig:phantomMaps20}  The top row shows the sensitivity maps generated for the reference method by dividing the image from each coil by the square root-sum-of-squares image.  The second and third row shows the sensitivity maps determined using the SAKE+L1 ESPIRiT and MCCS methods for the results presented in Fig. \ref{fig:phantomRecons20}. }
\end{figure*}

Figure \ref{fig:phantomMaps20} shows the sensitivity maps of the reference data as well as those estimated by SAKE+L1 ESPIRiT and MCCS.  The sensitivity maps of MCCS are much more similar to the reference maps than those of SAKE+L1 ESPIRiT.  Moreover, the sensitivity maps of SAKE+L1 ESPIRiT are not smooth enough to be consistent with the Biot-Savart law.

\begin{figure*}
  \centering{}
  \includegraphics[width=0.8\linewidth]{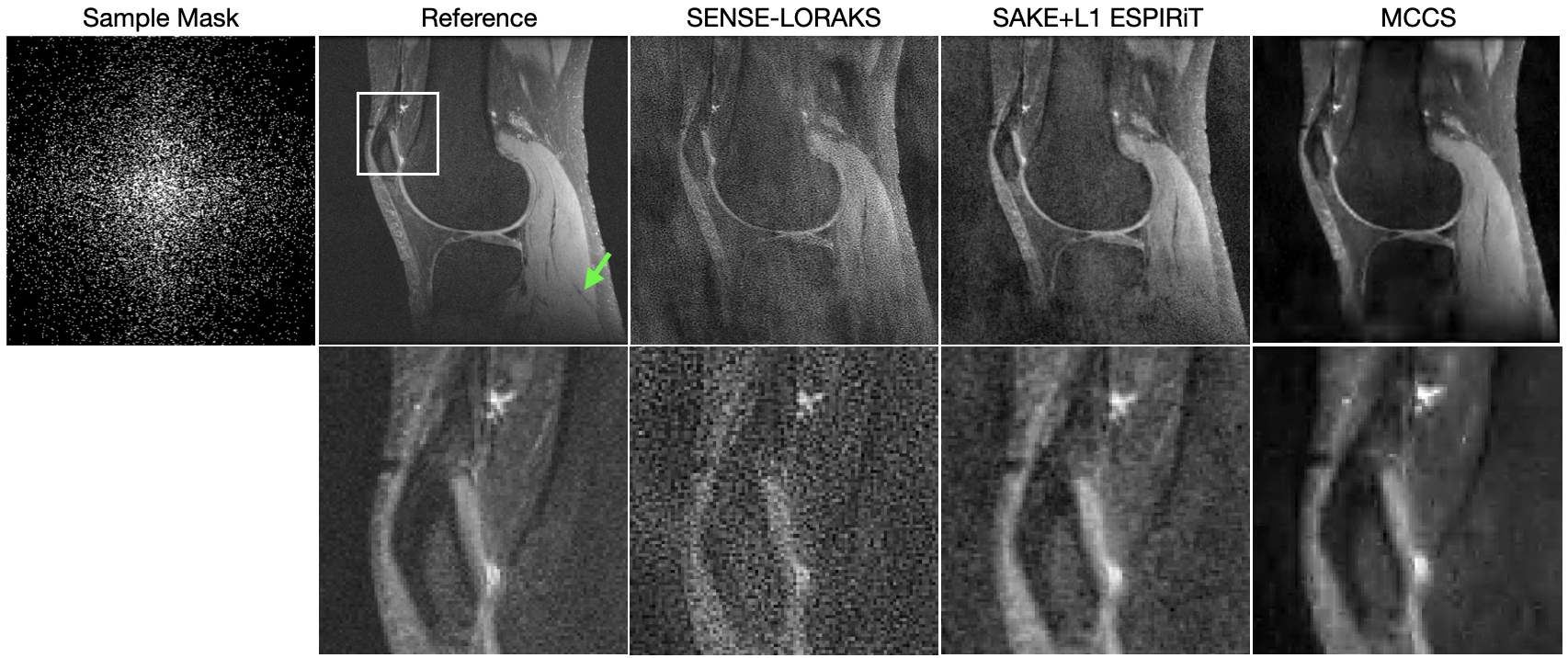}
  \captionsetup{width=.95\linewidth}
  \caption{ \label{fig:kneeRecons15}  The left column shows the sample mask used for this data (white points represent samples that were collected).  The top row shows the full field of view, the bottom row is zoomed into the white box overlaid on top of the reference image.  The third, fourth, and fifth columns present reconstructions using SENSE-LORAKS, SAKE+L1 ESPIRiT, and MCCS with the sample mask shown, which is 15\% of the data require for full sampling (that which satisfies the Nyquist-Shannon theorem).  The green arrow points to a detail that is not apparent in the SENSE-LORAKS or SAKE+L1 ESPIRiT reconstruction but is visible in the MCCS reconstruction.}
\end{figure*}

Figure \ref{fig:kneeRecons15} shows reconstructions of the knee using data from \url{mridata.org}.  The regularization parameter for SAKE+L1 ESPIRiT that provided the highest quality result was $2.5\cdot 10^{-3}$, which is the default supplied with the software.  Recall that the Biot-Savart simulation found a FWTM frequency of $5$ cycles per meter for this coils used in this dataset.  The regularization parameters used for MCCS were $(\lambda_x,\lambda_s)=(10^{-8},10^{-1})$; these were selected by hand to yield a high quality reconstruction.

At a gross level, one notes that the amount of noise present in the MCCS reconstruction is less than that of the SENSE-LORAKS and SAKE+L1 ESPIRiT reconstructions.  This is also true at the fine level, as seen in the second row.
The intensity of the muscle across the image is most uniform in the MCCS reconstruction.  Note that the muscle tissue in the posterior portion of the knee is more uniform in the MCCS reconstruction even than the reference reconstruction with fully sampled data.  The striation in the muscle indicated by the green arrow in the reference image is most prominent in the MCCS reconstruction.

\begin{figure*}
  \centering{}
  \includegraphics[width=0.8\linewidth]{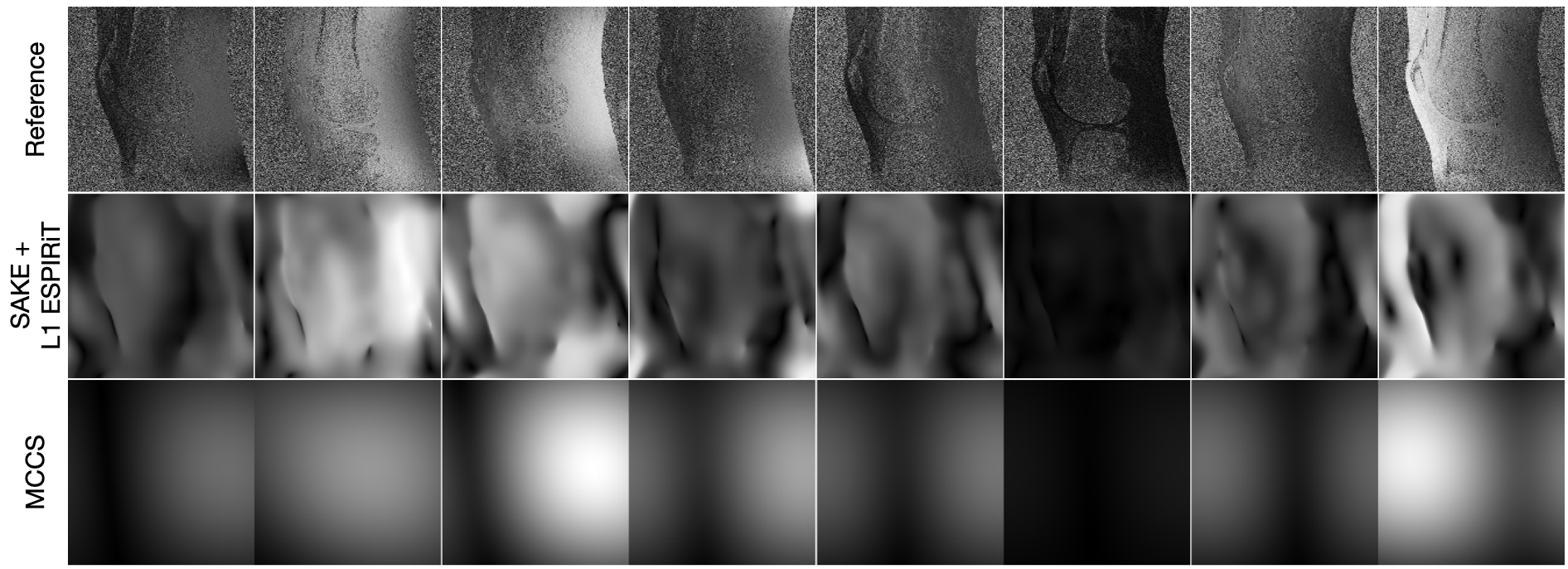}
  \captionsetup{width=.95\linewidth}
  \caption{ \label{fig:kneeMaps15}  The top row shows the sensitivity maps generated for the reference method by dividing the image from each coil by the square root-sum-of-squares image.  The second and third row shows the sensitivity maps determined using the SAKE+L1 ESPIRiT and MCCS methods for the results presented in Fig. \ref{fig:kneeRecons15}. }
\end{figure*}

Figure \ref{fig:kneeMaps15} shows the sensitivity maps created for the reconstruction of Fig. \ref{fig:kneeRecons15}.  The reference method uses 100\% of the data, while the SAKE+L1 ESPIRiT and MCCS methods generated these maps using 15\% of the data.  Again, the maps generated by the SAKE+L1 ESPIRiT algorithm have sharp edges that do not satisfy the Biot-Savart law.  However, the sensitivity maps generated by MCCS are smoother and are more consistent with the Biot-Savart law.  This partially explains why the muscle tissue is more uniform across the reconstruction in Fig. \ref{fig:kneeRecons15} with MCCS.

\begin{figure*}
  \centering{}
  \includegraphics[width=0.8\linewidth]{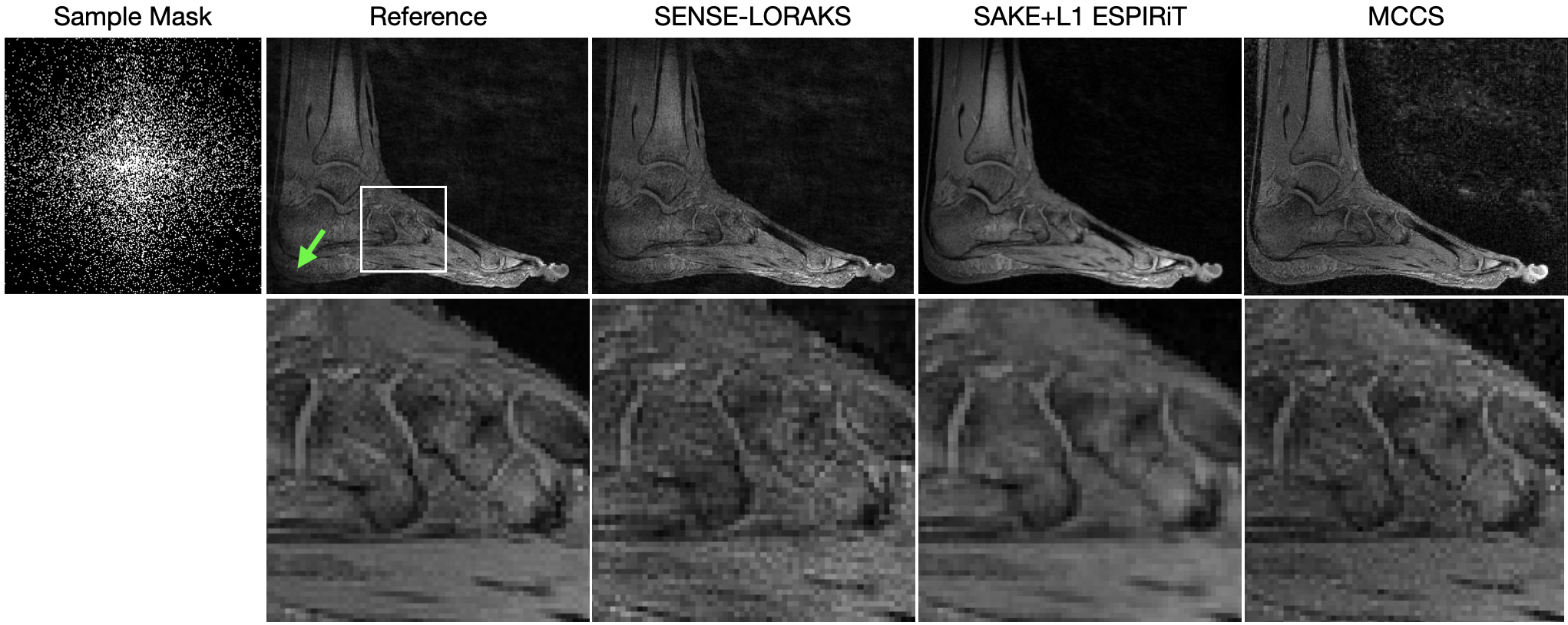}
  \captionsetup{width=.95\linewidth}
  \caption{ \label{fig:ankleRecons20}  The left column presents a reconstruction with 100\% of the samples for reference.  The other columns present reconstructions using SENSE-LORAKS, SAKE+L1 ESPIRiT, and MCCS with 15\% of the data require for full sampling (that which satisfies the Nyquist-Shannon theorem).  The green arrow points to the heel of the foot, which is most similar to surrounding tissue in the MCCS reconstruction (even more than the reference).  The second row zooms into the region shown in the white box of the reference image. }
\end{figure*}

Figure \ref{fig:ankleRecons20} shows reconstructions of the ankle using data collected from a $3$ Tesla scanner with an 3-element ankle coil.  The regularization parameter used for SAKE+L1 ESPIRiT was $2.5\cdot 10^{-3}$.  The regularization parameters used for MCCS were $\lambda_x,\lambda_s,\lambda_h)=(10^{-0},10^{0},10^9)$.

At a gross level, one notes that the amount of noise present in the MCCS reconstruction is less than that of the SENSE-LORAKS reconstruction.  The intensity throughout the anatomy is more uniform with MCCS than with the other reconstruction methods; this is highlighted by the more distinct outline of the heel, as indicated by the green arrow in the reference image.  Note that the tissue of the heel in the MCCS reconstruction is more similar to tissue of the same type in the rest of the ankle, even when compared to the reference image (where the heel is darker than it should be). When observing the zoomed in region in the second row, note the reduced noise of MCCS over SENSE-LORAKS and the increased detail of MCCS of SAKE+L1 ESPIRiT.

\begin{figure*}
  \centering{}
  \includegraphics[width=0.8\linewidth]{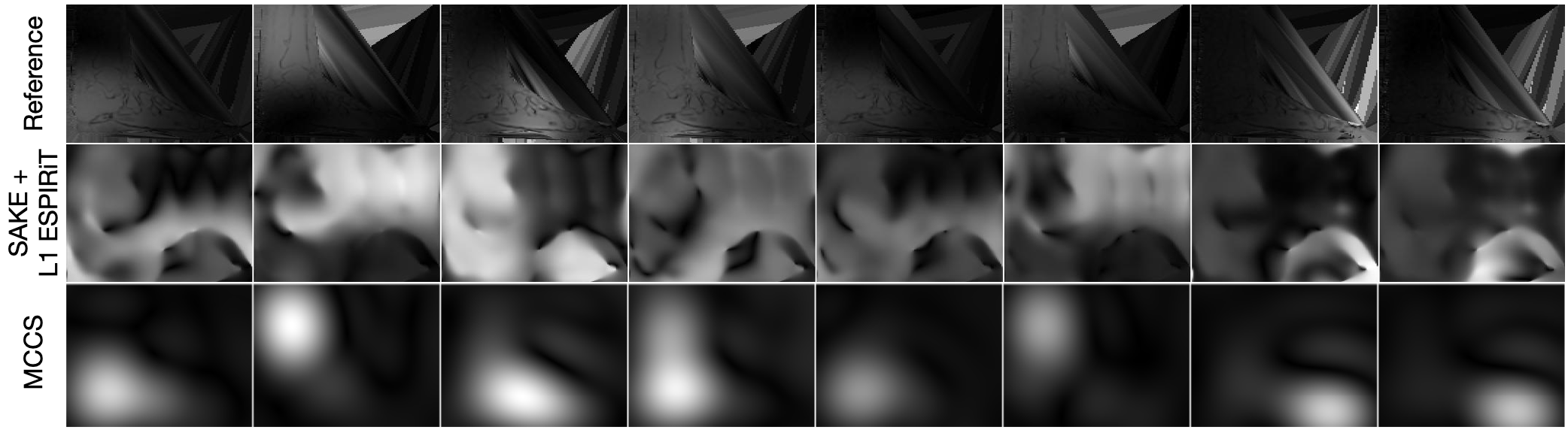}
  \captionsetup{width=.95\linewidth}
  \caption{ \label{fig:ankleMaps20}  The top row shows the sensitivity maps generated for the reference method by dividing the image from each coil by the square root-sum-of-squares image; the value for pixels with sufficiently small magnitude are determined with nearest neighbor interpolation.  The second and third row shows the sensitivity maps determined using the SAKE+L1 ESPIRiT and MCCS methods for the results presented in Fig. \ref{fig:ankleRecons20}. }
\end{figure*}

Figure \ref{fig:ankleMaps20} shows the sensitivity maps estimated with the reconstructions of Fig. \ref{fig:ankleRecons20}.  As before, the sensitivity maps estimated by SAKE+L1 ESPIRiT have edges that are inconsistent with the Biot-Savart law.  As before, the sensitivity maps estimated by MCCS are more smooth and more consistent with the Biot-Savart law.

\section{Discussion}

MCCS can generate results of higher quality than the other algorithms tested.  This comes at the expense of manually tuning four parameters and a much higher computational cost.

For MCCS to become clinically applicable, an effective method of automatically selecting the regularization parameters must be implemented.  For $\lambda_x$, an iterative reweighting scheme may be used \cite{candes2008enhancing}.  For the $\lambda_s$, if a sufficiently sized set of data of a particular anatomy existed, then they could be chosen with a multi-level minimization \cite{chung2017learning,ying2004tikhonov}.  Alternatively, a proper selection of regularization parameters may be possible by satisfying the Residual Whiteness Principle \cite{lanza2020residual}.  These possibilities are left for future investigations.

The MCCS algorithm is currently implemented in Matlab without parallelization.  It takes over an hour to generate a single image with a single core on a 2019 Mac Pro.
Future work will focus on increasing the runtime of MCCS; several algorithmic may do so.  For example, it can be altered to incorporate coil compression in order to reduce the number of computations \cite{zhang2013coil}.  (Coil compression can be achieved by constraining the rank of the $\mathcal{S}$ matrix rather than penalizing its nuclear norm.)  Additionally, preconditioning can be used to reduce the number of iterations required for each sub-problem \cite{pock2011diagonal,ong2019accelerating}.  In terms of its implementation, rather than using Matlab, the algorithm can be implemented in C and take advantage of GPU hardware for increased speed.  We expect these modification to yield images within a few minutes.

There may be several natural algorithmic extensions of MCCS to improve the quality further.  Currently, we are treating each two dimensional slice of the data independently and working in a hybrid space (the data is preprocessed with an inverse Fourier transform in the readout dimension, but not in the spatial dimensions).  However, the limited bandwidth constraint and small nuclear norm assumption for the sensitivity maps are valid in three-dimensions.  This could be taken into account during the optimization.  Additionally, the reconstruction algorithm can be adapted to accept data collected with non-Cartesian trajectories.  The difference with the method presented in this manuscript is that the product $\boldsymbol{D}\boldsymbol{F}$ would change to a non-uniform Discrete Fourier Transform, which will depend on the locations of the samples collected.  Future work can attempt to take advantage of linear predictability \cite{haldar2018linear}, a calibration region \cite{mcmanus2023dependence}, and the structure of the wavelet transform \cite{dwork2021utilizing,dwork2022utilizing} .  Finally, additional physics can be used to further constrain the solution with the use of Maxwell regularization \cite{francavilla2021maxwell}.

\section{Conclusion}

The work presented in this paper builds off a long history of innovations in MR image reconstruction.  The MCCS algorithm, presented in this manuscript, lends credence to the idea that the more physics is incorporated into the solution, the higher the quality of that solution will be.  In this work, we have shown that if the regularization parameters are chosen appropriately, then the reconstructions by MCCS are of higher quality than those of SENSE LORAKS or SAKE+L1 ESPIRiT.  These improvements come at the cost of significantly higher computational complexity and the need to manually select regularization parameters.


%

\appendix

\section{Table of notation}
\label{sec:notationTable}

Here we list the symbols used in the main manuscript and their meaning.
\begin{center}
\noindent\begin{tabularx}{\linewidth}{ |c|X| } 
 \hline
 $\odot$ & Hadamard product \\
 $\otimes$ & Kronecker product \\
 $\left(\cdot,\cdot,\ldots,\cdot\right)$ & vertical concatenation of elements \\
 $\|\cdot\|_1$ & $L_1$ norm \\
 $\|\cdot\|_{2,1}$ & $L_{2,1}$ norm \\
 $\|\cdot\|_{\ast}$ & nuclear norm \\
 $b$ & $\left(b^{(1)},b^{(2)},\ldots,b^{(C)}\right)$ \\ 
 $b^{(j)}$ & vector of data collected from the $j^{\text{th}}$ coil \\
 $\mathbb{C}$ & the set of complex numbers \\
 $\mathbb{C}^N$ & the set of vectors with $N$ elements over $\mathbb{C}$ \\
 $\text{Cov}(\cdot,\cdot)$ & creates a covariance matrix \\
 $\text{diag}$ & converts a vector into a diagonal matrix \\
 $D$ & diagonal sampling mask matrix \\
 $\boldsymbol{D}$ & $\diag(D, D, \cdots, D)$ \\
 $D_c$ & diagonal matrix with $1$ for every high frequency element and $0$ elsewhere \\
 $\boldsymbol{D}_c$ & $\diag(D_c,D_c,\cdots,D_c)$ \\
  $\eta^{(j)}$ & additive noise for the $j^{\text{th}}$ coil \\
 $\eta$ & $\left(\eta^{(1)},\eta^{(2)},\ldots,\eta^{(C)}\right)$ \\
 $C$ & Number of receiver coils \\
 $F$ & Discrete Fourier transform \\
 $\boldsymbol{F}$ & \text{diag}(F,F,\ldots,F) \\
 $\gamma$ & total variation regularization parameter \\
 $\cdot^H$ & Hermitian transpose \\
 $I_K$ & identity matrix of size $K\times K$ \\
 $K$ & the number of data elements collected from each coil \\
 $\lambda_x$ & regularization parameter that scales $\mathcal{R}_x$ \\
 $\lambda_s$ & regularization parameter that scales $\mathcal{R}_s$ \\
 $\tilde{\lambda}_s$ & high frequency regularization parameter \\
 $\boldsymbol{L}$ & left matrix of Cholesky decomposition of $\mathcal{N}^{-1}$ \\
 $M$ & number of rows of the image \\
 $\nabla$ & returns the gradient \\
 $N$ & number of columns of the image \\
 $\mathcal{N}$ & noise covariance matrix \\
 $\mathcal{R}_x$ & regularization function applied to image \\
 $\mathcal{R}_s$ & regularization function applied to sensitivity maps \\
 $\sigma^{(j)}$ & the sensitivity map of the $j^{\text{th}}$ coil in the continuous domain \\
 $s^{(j)}$ & the column extension of the $j^\text{th}$ sensitivity map \\
 $S^{(j)}$ & a diagonal matrix with diagonal elements equal to $s^{(j)}$ \\
 $\mathcal{S}$ & $[ s^{(1)} \, s^{(2)} \,  \cdots \, s^{(C)} ]$ \\
 $\boldsymbol{\mathcal{S}}$ & a block-column matrix equal to $\left(S^{(1)},S^{(2)},\ldots,S^{(C)}\right)$ \\
 $\cdot^\star$ & result from solving an optimization problem \\
 $W$ & wavelet transform \\
 $x$ & column extension of the image \\
 $X$ & a two-dimensional array of size $M\times N$ representing the image \\
 $\boldsymbol{X}$ & \diag(X,X,\ldots,X) \\
 \hline
\end{tabularx}
\end{center}

\section{Solving for the Sensitivity Maps}
\label{sec:pdhgForSenseMaps}

Primal-Dual Hybrid Gradient (PDHG) \index{Primal-Dual Hybrid Gradient,Chambolle-Pock} solves problems of the form: $\text{minimize} f(s) + g(As)$ where $f$ and $g$ are both closed convex proper (CC) with simple proximal operators and $A$ is a matrix \cite{chambolle2011first}.  By defining $f$, $g$ , and $A$ as follows, problem \eqref{eq:mccs_findMaps} can be solved with PDHG: $f(s) = \lambda_s \| \boldsymbol{S} \|_\ast$, $g(s) = (1/2)\| s^{(1)} - \boldsymbol{L}^H b\|_2^2 + \lambda_h / 2 || s^{(2)} ||_2^2 + \mathbb{I}_{\leq 1}(s^{(3)})$, and
\begin{equation*}
    A = \begin{bmatrix}
      \boldsymbol{D}\boldsymbol{F}\boldsymbol{L}^H\boldsymbol{X} & 0 & 0 \\
      0 & \sqrt{\lambda_h} \boldsymbol{D}_c \boldsymbol{F} & 0 \\
      0 & 0 & 0
      \end{bmatrix},
\end{equation*}
where $s^{(i)}$ is the $i^\text{th}$ portion of the vector $s$, and $\mathbb{I}_{\leq 1}$ is the indicator function that equals $0$ when all all components of the input vector are less than or equal to $1$ and equals $\infty$ otherwise.
The PDHG algorithm is as shown in Alg. \ref{alg:pdhg}.  For this algorithm, $\theta\in(0,2)$ is the relaxation parameter and $g^\ast$ is the conjugate function of $g$.
\begin{algorithm}[ht]
    \protect\caption{Primal-Dual Hybrid Gradient (PDHG)}
    \label{alg:pdhg}

    \textbf{Inputs:} $x^{(0)}$

    $z^{(0)} = 0$

    \textbf{For} $k = 1, 2, \ldots, K$

    \hspace{1em} $x^{(k+1)} = \prox_{\tau f}\left( x^{(k)} - \tau A^\ast z^{(k)} \right)$

    \hspace{1em} $\bar{x}^{(k+1)} = x^{(k+1)} + \theta\left( x^{(k+1)} - x^{(k)} \right)$
    
    \hspace{1em} $z^{(k+1)} = \prox_{\sigma g^\ast}\left( z^{(k)} + \sigma A \bar{x}^{(k+1)} \right)$

\end{algorithm}

The proximal operator of $f$ is a soft-threshold applied to the singular values of its input matrix.  The proximal operator of $g^\ast$ can be implemented using the separable sum rule where the proximal operator of the conjugate of the indicator function is a soft threshold of its input, and that of the conjugate of the scaled $g_1(s^{(1)}) = (1/2) \| s^{(1)} - \boldsymbol{L}^H b \|_2^2$ is
\begin{equation*}
  \prox_{\sigma g_1^\ast}(s^{1)} = \frac{1}{\sigma + 1}( s^{(1)} - \sigma \boldsymbol{L}^H b ).
\end{equation*}

\section{Solving for the images}
\label{sec:pogmForImage}
POGM solves problems of the form $\text{minimize}\hspace{0.5em} f(x) + g(x)$ where $f$ is differentiable and $g$ is CCP with a simple proximal operator \cite{fessler2019opt}.  By letting $f(x)=(1/2)\|\boldsymbol{D}\boldsymbol{F}\boldsymbol{L}^H\boldsymbol{S}x - \boldsymbol{L}^H b\|_2^2$ and $g(x)=\lambda_x\|W x\|_1$, we see that problem \eqref{eq:mccs_findImage} can be solved with the POGM method.
The POGM algorithm is shown in Alg. \ref{alg:pogm}, where $\nabla$ represents the gradient.  The proximal operator of $g$ is $\text{prox}_g(x) = W^\ast \tau\left( W x \right)$.

\begin{algorithm}[ht]
    \protect\caption{Proximal Optimal Gradient Method (POGM)}
    \label{alg:pogm}

    \textbf{Inputs:} $x^{(0)}$

    $\theta_0=1; w^{(0)} = x^{(0)}$

    \textbf{For} $k = 1, 2, \ldots, K$

    \hspace{1em} $\theta^{(k)} = 0.5 \left( 1 + \sqrt{ (4+4^{k>1}) \theta_{k-1}^2 + 1 } \right)$

    \hspace{1em} $\gamma = t * ( 2*\theta_{k-1} + \theta_k - 1 ) / \theta_k$

    \hspace{1em} $w^{(k)} = x^{(k-1}) - t * \nabla g( x^{(k-1)}$

    \hspace{1em} $z^{(k)} = w^{(k)} + ( \theta_{k-1} - 1 ) (w^{(k)} - w^{(k-1)} / \theta_{k} $
    
    \hspace{2em} $+ \theta_{k-1} ( w^{(k)} - x^{(k-1)} )/ \theta_{k}$
    
    \hspace{2em} $+ t ( \theta_{k-1} - 1 ) ( \gamma \theta_{k} ) / ( z^{(k)} - x^{(k-1)} )$

    \hspace{1em} $x^{(k)} = \prox_{t g}\left( z^{(k)} \right)$

    \textbf{End For}
\end{algorithm}

\section*{Acknowledgments}
ND would like to thank the Quantitative Biosciences Institute at UCSF and the American Heart Association as funding sources for this work.
PL and ND would like to thank the National Institute of Health for grant R01HL136965 as a funding source of this work.
The authors would like to thank Mary Frost, Kim Okomato, and Heather Daniel for their assistance in collecting data.
The authors would like to thank Michael Ohliger for useful discussions regarding electromagnetics and receiver coils.


\end{document}